\documentclass[superscriptaddress,prd,aps,twocolumn,floats,nofootinbib,amsfonts,amssymb,preprintnumbers,showpacs]{revtex4}
\usepackage{mathrsfs,graphicx,bm,amsmath}
\begin{document}

\title{\bf Gauge-Invariant Initial Conditions and Early Time Perturbations in Quintessence Universes}
\author{Michael Doran}
\affiliation{Department of Physics and Astronomy, 6127 Wilder Laboratory, Hanover, New Hampshire 03755, USA}
\author{Christian M. M{\"{u}}ller}
\affiliation{Institut f{\"{u}}r Theoretische Physik, Philosophenweg 16, 69120 Heidelberg, Germany}
\author{Gregor Sch{\"{a}}fer}
\affiliation{Institut f{\"{u}}r Theoretische Physik, Philosophenweg 16, 69120 Heidelberg, Germany}
\author{Christof Wetterich}
\affiliation{Institut f{\"{u}}r Theoretische Physik, Philosophenweg 16, 69120 Heidelberg, Germany}
\date{April 14, 2003}

\newcommand{\norm}[1]{\mathcal{N}_{\rm #1}}

\def\dblone{\hbox{$1\hskip -1.2pt\vrule depth 0pt height 1.6ex width 0.7pt
            \vrule depth 0pt height 0.3pt width 0.12em$}} 
\newcommand{\Psiflex}{\Phi}
\newcommand{\Phiflex}{\Psi}

\newcommand{\isocdm}{CDM isocurvature}
\newcommand{\isobar}{baryon isocurvature}
\newcommand{\isonu}{neutrino isocurvature}
\newcommand{\Vphi}{V_{q}}
\newcommand{\Vphitilde}{\tilde{V}_{q}}
\newcommand{\Vphitildedot}{\dot{\tilde{V}}_{q}}
\newcommand{\Vphidot}{{\dot V}_{q}}
\newcommand{\Deltaphi}{\Delta_{q}}
\newcommand{\Deltaphidot}{{\dot\Delta}_{q}}
\newcommand{\wphi}{w_{q}}
\newcommand{\wq}{\wphi}
\newcommand{\Omegaphi}{\Omega_{q}}
\newcommand{\Mp}{{\rm M}_{\bar P}}

\newcommand{\Dg}[1]{\Delta_{#1}}
\newcommand{\rd}{\textrm{d}}

\newcommand{\hubble}{\mathcal{H}}

\newcommand{\dotphi}{\dot{\bar\varphi}}

\newcommand{\adota}{\frac{\dot a}{a}}

\newcommand{\ee}{\begin{equation}}
  \newcommand{\eee}{\end{equation}}
\newcommand{\ea}{\begin{eqnarray}}
  \newcommand{\eea}{\end{eqnarray}}

\newcommand{\ddotphi}{\ddot{\bar\varphi}}

\newcommand{\mpc}{\textrm{Mpc}}
\newcommand{\cpm}{\textrm{Mpc}\ensuremath{^{-1}}}
\newcommand{\cmbeasy}{CMBEASY\ }

\newcommand{\ome}[2][]{\ensuremath{
    \Omega^{#1}_{#2}}}
\newcommand{\omebar}[2][]{\ensuremath{
    \bar{\Omega}_{#1}^{#2}}}
\newcommand{\omc}{\ome[0]{c}}
\newcommand{\omb}{\ome[0]{b}}
\newcommand{\omq}{\ome[0]{\varphi}}
\newcommand{\omca}{\Omega_{c}}
\newcommand{\omba}{\Omega_{b}}
\newcommand{\omqa}{\Omega_{q}}
\newcommand{\omga}{\Omega_{\gamma}}
\newcommand{\omna}{\Omega_{\nu}}
\newcommand{\omma}{\Omega_{m}}
\begin{abstract}
We present a systematic treatment of the initial conditions
and evolution of cosmological perturbations in a universe
containing photons, baryons, neutrinos, cold dark matter, and a
scalar quintessence field.
By formulating the evolution in terms of a differential equation involving a matrix acting on a vector
comprised of the perturbation variables, we can use the familiar language of eigenvalues and
eigenvectors. As the largest eigenvalue of the evolution matrix is fourfold degenerate, it follows
that there are four dominant modes with non-diverging gravitational potential at early times, corresponding 
to 
adiabatic, cold dark matter isocurvature, baryon isocurvature and neutrino isocurvature perturbations. 
We conclude
that quintessence does not lead to an additional independent mode.
\end{abstract}

\preprint{HD--THEP--03--19}

\maketitle


\section{Introduction}

The advent of high precision data \cite{Spergel:2003cb} of the cosmic
microwave background (CMB) anisotropies permits detailed tests of the
composition and shape of the primordial density fluctuations. The most
popular models of inflationary cosmology predict adiabatic
fluctuations \cite{Mukhanov:xt,Bardeen:qw,Liddle:cg}. More elaborate models lead to an
admixture of adiabatic and isocurvature fluctuations
\cite{Lyth:2002my,Bartolo:2002vf}.  The time evolution of adiabatic
and non-adiabatic fluctuations is well understood for a universe
composed of radiation, baryons, cold dark matter (CDM) and neutrinos
\cite{Ma:1995ey}.  In the context of quintessence
\cite{Wetterich:fm,Ratra:1987rm,Caldwell:1997ii}, the behaviour of the
field fluctuation has been studied in several works
\cite{Viana:1997mt,dave_phd,Malquarti:2002iu,Abramo:2001mv,Kawasaki:2001nx}. Initial
conditions have been proposed in \cite{Perrotta:1998vf} for the case of
negligible quintessence contribution in the early universe. We present here a systematic treatment of 
initial conditions for 
quintessence models which differs from that of \cite{Perrotta:1998vf} in approach and interpretation.

Our basic setting assumes that small deviations from homogeneity are generated during a very early
stage of the big bang, typically an inflationary epoch. During the following radiation dominated
period the wavelength of the relevant fluctuations is far outside the horizon. Apart from this, we will not use any further 
constraint on the primordial fluctuations. Only the spectra
of a certain number of ``dominant'' modes can possibly influence events such as emission of the CMB
and its anisotropies since the other modes decay. The information about these dominant modes therefore constitutes the 
initial conditions
for practical purposes. Primordial information beyond the dominant modes is effectively lost and 
not observable. The detailed time of specification of the initial conditions is therefore irrelevant as long 
as it is much shorter than the time of matter-radiation equality. 

During the period relevant  for the discussion of the initial conditions the universe is
radiation dominated. However, our approach allows for the presence of scalar
fields which evolve like radiation at early times or are subdominant.
Consequently, our results hold for a wide class of quintessence
models, including those with non-negligible $\omqa$ at early times
\cite{Caldwell:2003vp}. In fact, we only use a ``tracking'' property
\cite{Steinhardt:nw} for the background of homogenous quintessence,
namely that its equation of state $w_q=p_q/\rho_q$ is almost constant
and determined only by the energy densities of the radiation and
matter components. The parameters $w_q$ and $\omqa = 1 - \omma - \omna
- \omga$ will therefore be the only parameters of the quintessence
model that influence the early time evolution of small fluctuations. This
makes our analysis model independent to a large extent.

We will formulate the evolution equations for the perturbation
variables as a first order differential matrix equation:
\begin{equation}\label{eqn::matrix}
\frac{\textrm{d}}{\textrm{d} \,\ln x} \bm{U} = A(x) \bm{U},
\end{equation}
where the vector $\bm{U}$ contains all perturbation variables and
the matrix $A(x)$ encodes the evolution equations.
In doing so,
we relate the problem of finding initial conditions and dominant modes
to the familiar language of eigenvalues and eigenvectors. This
formulation makes ``mode-accounting'' transparent by counting the
degeneracy of the largest eigenvalue.
We find four dominant modes that remain regular
at early times. For physical reasons, we choose a basis 
using 
adiabatic, \isocdm, \isobar\ and \isonu \ initial conditions.
As we will show, adiabaticity between CDM, baryons and photons
implies adiabaticity of quintessence. There is therefore 
no pure quintessence isocurvature mode. 
In addition, using the matrix formulation reveals facets  of the modes
that otherwise remain obscured.

In order to avoid the appearance of gauge modes, we 
will use the gauge-invariant formalism 
 \cite{Bardeen:kt,Kodama:bj,Mukhanov:1990me,Durrer:2001gq}. In contrast to 
earlier work, we find it more appropriate to specify the initial conditions
and time evolution of  the quintessence field in terms of the gauge-invariant density contrast and velocity, thus unifying the language 
for all species. As anticipated, the quintessence density perturbation
remains constant at super-horizon scales for adiabatic initial conditions.
In contrast to this, the field fluctuation follows a simple power law in
conformal time that only depends on the quintessence equation of state.

We will proceed as follows: in section \ref{section_gi} we give the
gauge-invariant perturbation equations for a radiation-dominated
universe containing radiation, cold dark matter, neutrinos, baryons in
the tight coupling limit and tracking quintessence. We express
the evolution in matrix form in \ref{sec::matrix}. 
In section
\ref{sec::class}, we classify the modes and determine them in
sections \ref{sec::adia}, \ref{sec::isonu} and \ref{sec::isocdm}. To illustrate the
effect of non-adiabatic contributions to the CMB spectrum, we plot a
few spectra for different initial conditions in section
\ref{section_example}. A summary of our findings is given in section
\ref{section_conclusion}. In Appendix \ref{section::gipe}, we derive the perturbation
equations used in detail, while Appendices \ref{early} and \ref{sec::extend} discuss
supplementary issues.

\section{The Perturbation Equations} \label{section_gi}
In the following we  adopt
the  gauge-invariant approach as devised by Bardeen \cite{Bardeen:kt}.
It is not difficult to obtain the initial conditions in any
gauge from the corresponding gauge-invariant quantities given
here. 
In Appendix A, we summarize the definitions of the perturbation
variables and sketch the derivation of the evolution equations. It
turns out that the evolution is best described as a function of $x
\equiv k \tau$, where $\tau$ is the conformal time and $k$ the comoving
wavenumber of the mode.
 We  assume
that at early times, the universe expands as if radiation dominated. This assumption is well
justified for small $\omqa$ at early times, as well as for potentials that are essentially
exponentials at the time of interest, regardless of $\omqa$. The assumption is 
certainly not justified for models in which quintessence is dominating the universe
at early times with equation of state $\wq \neq 1/3$. For such (slightly exotic) models, the following
steps would need to be modified.

\subsection{Full Set of Equations}
Assuming tracking quintessence we obtain the 
following set of equations (for a derivation, see Appendix \ref{section::gipe}):
\begin{eqnarray}\label{eq_dcstrich}
\Dg{c}'& =& - x^2 \tilde V_c, \\  \label{eq_vcstrich}
\tilde V_c'& =& -2 \tilde V_c + \Phiflex, \\ \label{eq_dgammastrich}
\Delta_{\gamma}'& =& -\frac{4}{3} x^2 \tilde V_{\gamma}, \\ \label{grand_equation_aa}
\tilde V_{\gamma}' &=&  \frac{1}{4} \Delta_{\gamma}- \tilde V_{\gamma}  + \Omega_{\nu} \tilde{\Pi}_{\nu} + 2\Phiflex ,  \\
\Dg{b}' &=& -x^2 \tilde V_{\gamma} \label{eq_dgbaryonstrich}, \\
\Delta_{\nu}'& =& -\frac{4}{3} x^2 \tilde V_{\nu}, \label{eq_dnustrich}
\end{eqnarray}
\begin{eqnarray}
\label{grand_equation_bb}
\tilde V_{\nu}'& =&  \frac{1}{4} \Delta_{\nu} - \tilde V_{\nu} -\frac{1}{6}x^2 \tilde \Pi_{\nu} + \Omega_{\nu} \tilde{\Pi}_{\nu} + 2\Phiflex  ,\\
\tilde \Pi_{\nu}'&=&\frac{8}{5}\tilde V_{\nu}-2 \tilde \Pi_{\nu}, \\
 \label{eq_q1} 
\Deltaphi' &=& 3 (\wq-1)\bigg[  \Deltaphi + 3(1+\wq)\left\{ \Phiflex + \Omega_{\nu} \tilde{\Pi}_{\nu}\right \} \nonumber \\
 &&+ \left \{3 - \frac{x^2}{3(\wq -1)} \right \}\, (1+\wq) \Vphitilde \bigg ],\\
 \label{eq_q2}
\Vphitilde'&=& 3\Omega_{\nu} \tilde{\Pi}_{\nu}  +\frac{\Deltaphi}{1+\wphi} + \Vphitilde   +4 \Phiflex, 
\end{eqnarray}
with the gauge-invariant Newtonian potential $\Phiflex$ given by 
\begin{equation} \label{eq_poisson_sum}
\Phiflex=- \frac{ \sum\limits_{\alpha =c,b,\gamma,\nu,q} \Omega_{\alpha} ( \Delta_{\alpha} +3(1+w_{\alpha})
\tilde{V}_{\alpha})}{ \sum\limits_{\alpha =c,b,\gamma,\nu,q} 
 3(1+w_{\alpha})\Omega_{\alpha} + \frac{2 x^{2}}{3}} -\Omega_{\nu} \tilde{\Pi}_{\nu}.
\end{equation}
We denote the derivative $\textrm{d}/\textrm{d} \ln x$ with a prime.
The gauge-invariant energy density contrasts  $\Delta_{\alpha}$, the velocities
$\tilde{V}_{\alpha}$ and the shear $\tilde \Pi_{\nu}$ are the ones
found in the literature \cite{Bardeen:kt,Kodama:bj,Durrer:2001gq}, except that 
we factor out powers of $x$
from the velocity and shear defining $\tilde{V} \equiv V/x$ and $\tilde{\Pi}_{\nu} \equiv x^{-2} \Pi_{\nu}$.
This factoring out leads to the particularly simple form of the system of equations for $x  \ll 1$ 
(see also Appendix \ref{section::gipe}).
It does, however, exclude modes with diverging $\Phiflex$ at early times such as
a neutrino velocity mode \cite{Rebhan:1994zw}.
The index $\alpha$ runs over the five
species in our equations, namely cold dark matter, baryons, photons, neutrinos
and quintessence, denoted with the subscript $q$. We assume tight
coupling between photons and baryons. The equation of state
$w=\bar{p}/\bar{\rho}$ takes on the values $w_{c}=w_b=0$,
$w_{\gamma}=w_{\nu}=1/3$ and $\wphi$ is left as a free parameter.
Equations
(\ref{eq_dcstrich}), (\ref{eq_dgammastrich}),   (\ref{eq_dgbaryonstrich}) and (\ref{eq_dnustrich})
can be regarded as continuity relations between the density
fluctuations and the velocity.  We obtain equations (\ref{eq_q1}) and
(\ref{eq_q2}) from the perturbed Klein-Gordon equation of the
quintessence scalar field expressed in terms of $\Deltaphi$ and
$\Vphi$, the energy density and velocity perturbations as defined in Appendix \ref{section::gipe}.


\subsection{Matrix Formulation and Dominant Modes}\label{sec::matrix}

Conceptually, it is convenient to note that 
the above set of equations can be concisely written
in matrix form according to Equation (\ref{eqn::matrix}) where the perturbation vector is defined as
\begin{equation}
\bm{U}^{T} \equiv (\Dg{c},\,\tilde{V}_{c},\,\Delta_{\gamma},\,\tilde{V}_{\gamma},
\,\Dg{b},\,\Delta_{\nu},\,\tilde{V}_{\nu},\,\tilde{\Pi}_{\nu},\,\Deltaphi,\,\Vphitilde).
\end{equation}
The matrix $A(x)$ can easily be read off from equations (\ref{eq_dcstrich})-(\ref{eq_q2}).
This enables us to discuss the problem of 
specifying initial conditions in a systematic way. 

The initial conditions are specified for modes well outside the horizon, i.e.
$x \ll 1$.  In this case,  the r.h.s. of equations
(\ref{eq_dcstrich}),\ (\ref{eq_dgammastrich}), (\ref{eq_dgbaryonstrich}) and (\ref{eq_dnustrich})
can be neglected, provided $\tilde{V}_{\alpha}$ does not diverge $
\propto x^{-2}$ or faster for $x^2\to0$. The evolution matrix $A(x)$ loses
any explicit $x$ dependence for $x^2\to0$. Yet, it still depends on $x$ via terms
involving $\omca, \omba$ and $\omqa$. By our assumptions on quintessence,
the term involving $\omqa$ is either a constant (for $\wq = 1/3$) or negligible
 (yet, in Appendix \ref{sec::extend}, we extend the treatment to include models with 
considerable $\omqa$ and $\wq \neq 1/3$.) In both cases $\omqa$ can be approximated by a 
constant ($\omqa = 0$ for $ \wq < 1/3$) and 
$\omca$, $\omba$ vanish $\propto x$. In leading order, the matrix $A$ becomes therefore
 $x$-independent for very early times.
In fact, the general solution to Equation (\ref{eqn::matrix}) in the (ideal) case of 
a truly constant $A$ would be 
\ee\label{eigenvector_decomposition}
\bm{U}(x) = \sum\limits_i c_{i} \left( \frac{x}{x_0}\right)^{\lambda_i}\bm{U}^{(i)},
\eee
where $\bm{U}^{(i)}$ are the eigenvectors of $A$ with eigenvalue $\lambda_i$ and the
time independent coefficients $c_i$ specify the initial contribution of $\bm{U}^{(i)}$ towards a 
general perturbation $\bm{U}$.
As time progresses, components corresponding to the largest eigenvalues $\lambda_i$
will dominate. Compared to these ``dominant'' modes, initial contributions  
in the direction of eigenvectors  $\bm{U}^{(i)}$ with smaller ${\rm Re}(\lambda_i)$ 
decay. It therefore suffices to specify the initial contribution $c_i$ for
the dominant modes, if one is not interested in very early time behaviour shortly
after inflation.
In our case,
the characteristic polynomial of $A(x)$ indeed has a fourfold degenerate
eigenvalue $\lambda = 0$ in the limit $x^2 \to 0$, independent of $\omca$, $\omba$ and 
$\omqa$.\footnote{For $w_q=1$ we find another eigenvalue with $\lambda=0$. We will ignore this special 
case in what follows.}
While it is not feasible to obtain the remaining six eigenvalues by analytic means, we have
checked numerically for a wide range of
$\Omega_{\gamma}$, $\Omega_{\nu}$, $\omba$, $\omca$, 
  $\omqa$ and $\wphi$ that the remaining eigenvalues have indeed negative real parts and contributions
from the corresponding eigenvectors towards a general perturbation $\bm{U}$ will therefore
decay according to Equation (\ref{eigenvector_decomposition}). We can improve the analytic 
description of the dominant modes by taking
corrections $\propto x$ into account.

As $\omca \propto \omba \propto x$, it is appropriate to split $A(x)$ according to the scaling with $x$,
\ee \label{eq_A}
A = A_0 + x \, A_1,
\eee
where $A_0$ and $A_1$ are constant and $x \,A_1$ contains the small, time-dependent corrections
from terms involving $\omca$ and  $\omba$. 
We may also write\footnote{This form is not an ansatz, but dictated by Equation (\ref{eqn::matrix}), once
the dependence of $A(x)$ on $x$ is given.}  the eigenvectors as a series in $x$,
\ee \label{eq_U}
\bm{U}= \bm{U}_0 + x\,\bm{U}_1.
\eee
Inserting Equations (\ref{eq_A})-(\ref{eq_U}) in Equation (\ref{eqn::matrix}), we get
\ee\label{eqn::a0}
A_0\, \bm{U}_0 = 0,
\eee
and 
\ee\label{eqn::a1}
\bm{U}_1 = - (A_0 - \openone)^{-1} A_1 \bm{U}_0.
\eee
Equation (\ref{eqn::a1}) is easy to solve, once $\bm{U}_0$ has been determined (we 
discuss the possibility of a vanishing $\bm{U}_0$ in Appendix \ref{sec::extend}). 
We see from eqn. (\ref{eqn::a0})
that to constant order the solutions of eqn. (\ref{eqn::matrix}) are indeed given 
by eigenvectors to the eigenvalue $\lambda=0$. 
We should emphasize that the vectors $\mathbf U_0$ do not evolve in time if their corresponding 
eigenvalues are $\lambda=0$. Thus, the perturbations remain constant in the super-horizon 
regime during radiation  domination in this approximation. If we include the next-to-leading 
order contribution to $\mathbf U$, the eigenvectors do evolve and we can no longer apply 
eq. (\ref{eigenvector_decomposition}). These corrections are, however, small as long as we 
are deep in the radiation dominated era due to the small contributions of baryons, radiation 
and quintessence during this era. Given a set of initial conditions in the form of coefficients 
for the four dominating modes at $z_{initial}$ we can find the perturbations at some later time 
(provided the modes are still super-horizon sized and we have radiation domination). In leading 
order, the coefficients will remain the same while in next-to-leading order we can use the 
evolution of $\mathbf U$ to compute the coefficients for $z< z_{initial}$. If initial conditions 
are specified with accuracy of next-to-leading order one therefore has to specify $z_{initial}$ 
as well. In leading order this is unecessary for $z$ in a wide range long before last scattering.


\subsection{Constraint Equations to Leading Order}
Equation (\ref{eqn::a0}) is equivalent to setting 
the l.h.s. of Equations (\ref{eq_dcstrich})-(\ref{eq_q2}) 
equal to zero and using $\omca=\omba=x^2=0$.
Then Equations (\ref{eq_dcstrich}),
(\ref{eq_dgammastrich}), (\ref{eq_dgbaryonstrich})  and (\ref{eq_dnustrich}) are automatically
satisfied (provided $\tilde{V}_{\alpha}$ does not diverge 
$\propto x^{-2}$ or faster), and  Equations (\ref{eq_vcstrich}),(\ref{grand_equation_aa}),(\ref{grand_equation_bb})-(\ref{eq_q2}) 
yield non-trivial constraints for the components of $\bm{U}_0$:
\begin{eqnarray} \label{constraint_equation_start}
2 \tilde V_c-\Phiflex &=& 0, \\
1/4 \Delta_{\gamma} - \tilde V_{\gamma}+\Omega_{\nu}\tilde \Pi_{\nu}+2\Phiflex &=& 0, \\
1/4 \Delta_{\nu} - \tilde V_{\nu}+ \Omega_{\nu}\tilde \Pi_{\nu} +2\Phiflex &=& 0, \\
8/5 \tilde V_{\nu}- 2\tilde \Pi_{\nu} &= & 0, \\  \label{constraint_equation_onebeforeend}
3 \Omega_{\nu}\tilde \Pi_{\nu} +\Deltaphi /(1+\wphi)+ 3 \Vphitilde + 3 \Phiflex &= & 0, \\ \label{constraint_equation_end}
3\Omega_{\nu}\tilde \Pi_{\nu}+\Deltaphi/(1+\wphi)+ \Vphitilde +4\Phiflex&=& 0. 
\end{eqnarray}
In the above, all quantities are considered only to constant order. 
(we have omitted the subscript '$0$' for notational convenience.)
In particular, there is no contribution of CDM and baryons  to $\Phiflex$ 
at constant order. 
Note that, apart from $\wphi$, no model-specific parameters occur in
any of these equations so the modes will be independent of the type of
quintessence as long as the scalar field is in a regime with
approximately constant $\wphi$. We note that for $\wphi$ substantially
smaller than $1/3$ the quintessence fraction $\Omegaphi$ changes with
time. By the assumption that the universe expands as if radiation dominated,
the quintessence contribution would however be small in this case and its contribution
to $\Phiflex$ can be neglected (see Appendix \ref{sec::extend} for an extended discussion).

We mention that for $w_q=1/3$, quintessence evolves the same way as radiation, 
therefore $\Omega_q$ does not change in this case. If $w_q=-1/3$, quintessence 
has the same influence on the scale factor $a$ as a curvature term in an open 
universe. However, the geometry is still flat and one can distinguish an open 
universe from this quintessence model by measuring the position of the first acoustic peak in the CMB.


\section{The modes in detail} \label{section_modes}

\subsection{Classifying the modes}\label{sec::class}

While any basis for the subspace spanned by the eigenvectors with
eigenvalue 
$\lambda=0$ can be used to specify the initial conditions, it is 
still worthwhile to use a basis that is physically meaningful.
Following the existing literature, we use the 
gauge-invariant entropy perturbation \cite{Kodama:bj}
\begin{equation}
S_{\alpha:\beta}= \frac{\Delta_{\alpha}}{1+w_{\alpha}}-\frac{\Delta_{\beta}}{1+w_{\beta}},
\end{equation}
between two species $\alpha$ and $\beta$, 
as well as the  gauge-invariant curvature perturbation on
hyper-surfaces of uniform energy density of species $\alpha$
\cite{Bardeen:qw,bardeen_isocurvature_book,Lyth:2002my,Wands:2000dp}
\begin{equation}
\zeta_{\alpha}= \left(H_L+\frac{1}{3} H_T\right)+\frac{\delta \rho_{\alpha} }{3(1+w_{\alpha})\bar \rho_{\alpha}},
\end{equation} 
in order to classify the physical modes.
On slices of uniform total energy density, the curvature perturbation is correspondingly
\begin{equation}
\zeta_{tot}=  \left(H_L+\frac{1}{3} H_T\right)+\frac{\sum_{\alpha} 
\delta \rho_{\alpha}}{\sum_{\alpha} 3(1+w_{\alpha}) \bar \rho_{\alpha}}.
\end{equation}
In our variables, these expressions take on the manifestly gauge-invariant form
\begin{equation}
\zeta_{\alpha}=\frac{\Delta_{\alpha}}{3(1+w_{\alpha})} \ , \ \ \  \zeta_{tot}=\frac{\sum_{\alpha} \Delta_{\alpha} \Omega_{\alpha}}{\sum_{\alpha} 3(1+w_{\alpha})\Omega_{\alpha}}.
\end{equation}
If $\zeta_{tot}=0$, energy density perturbations do not generate curvature.  It is therefore clear that such a perturbation is a perturbation in the local equation of state.
One should note that the definition of $\zeta_{tot}$ is different from that of \cite{Mukhanov:1990me}:
\begin{equation}
\zeta_{MFB}=\frac{2}{3}\frac{\hubble ^{-1}\dot \Phiflex+\Phiflex}{(1+w)}+\Phiflex.
\end{equation}
However, one may verify that this quantity coincides with $\zeta_{tot}$ in the super-horizon limit for a flat universe \cite{martin}.


\subsection{The Adiabatic Mode}\label{sec::adia}

The first (rather intuitive) perturbations one would try to find are adiabatic
perturbations, which are specified by the adiabaticity conditions
$S_{\alpha:\beta}=0$ for all pairs of components. In our case, this
results in eleven constraints\footnote{Without requiring quintessence to be adiabatic, we  have six  constraints from 
equations (\ref{constraint_equation_start})-(\ref{constraint_equation_end}) plus three constraints 
from eq. (\ref{adiabatic_conditions}) plus one constraint from the overall normalization, which 
is fixed by choosing a specific value for $\Delta_{\gamma}$.} for the ten components of $\bm{U}_0$. It
is a priori not clear that this has a solution so we will not include
quintessence in the adiabaticity requirement. Requiring adiabaticity
between CDM, baryons, neutrinos and radiation,
\begin{equation}\label{adiabatic_conditions}
\Delta_{\nu}=\Delta_{\gamma}=\frac{4}{3} \Dg{c} = \frac{4}{3} \Dg{b},
\end{equation}
and using the six constraint Equations (\ref{constraint_equation_start})-(\ref{constraint_equation_end}), we obtain
\begin{equation}\label{adiabatic_mode}
\left(
\begin{array}{c}
\Dg{c} \\
\tilde V_c\\
\Delta_{\gamma} \\
\tilde V_{\gamma} \\
\Dg{b} \\
\Delta_{\nu} \\
\tilde V_{\nu} \\
\tilde \Pi_{\nu} \\
\Deltaphi \\
\Vphitilde
\end{array}
\right)_{\mbox{adiabatic}} = C
\left(
\begin{array}{c}
3/4 \\
(-5/4)\, \mathcal{ P}\\
1 \\
(-5/4)\, \mathcal{P}\\
3/4 \\
1 \\
(- 5/4)\, \mathcal{P} \\
-\mathcal{P} \\
 3 (1+ \wphi) /4 \\
(-5/4)\, \mathcal{P}
\end{array}
\right),
\end{equation}
where $\mathcal{P}= (15 + 4 \Omega_{\nu})^{-1}$ and $C$ is an arbitrary  constant. From $\Delta_q/\Delta_{\gamma}=3 ( 1+\wq)/4$ 
we conclude that quintessence is automatically adiabatic if CDM, baryons,
neutrinos and radiation are adiabatic, independent of the quintessence
model for  as long as we are in the tracking regime.
As all components are non-vanishing, we do not quote the next to leading order
contributions from $x\,\bm{U}_1$.


\subsection{Neutrino Isocurvature}\label{sec::isonu}
Having found the adiabatic vector, one could specify three additional
linearly independent vectors satisfying the constraint Equations
(\ref{constraint_equation_start})-(\ref{constraint_equation_end}).
This would  complete the basis.  It is, however, appropriate to choose modes that
may be generated by physical processes.
These modes are in general not orthogonal but span the 
eigenspace of $\lambda=0$. 
Modes that may be generated by physical
processes are  isocurvature modes. A given mode is an
isocurvature mode, if the gauge-invariant curvature perturbation
$\zeta_{tot}$ vanishes, i.e. $\zeta_{tot}=0$. In order to distinguish different
isocurvature modes from one another, we require that the other species are adiabatic with respect to each other, i. e.
$S_{\alpha:\beta}=0$ except for quintessence and  one species $\sigma$, which has
non-vanishing $S_{\sigma:\gamma}$. 

Let us first consider the  neutrino isocurvature mode. For this, we
require that CDM, baryons  and radiation are adiabatic, while $S_{\nu:\gamma} \neq 0$
and that the gauge-invariant curvature perturbation vanishes:
\begin{equation}
\zeta_{tot} = 0, \ \ \ \Dg{c}= \Dg{b} = \frac{3}{4} \Delta_{\gamma}.
\end{equation}
Using this and Equations (\ref{constraint_equation_start})-(\ref{constraint_equation_end}) leads to
\begin{equation}
\left(
\begin{array}{c}
\Dg{c} \\
\tilde V_c\\
\Delta_{\gamma} \\
\tilde V_{\gamma} \\
\Dg{b} \\
\Delta_{\nu} \\
\tilde V_{\nu} \\
\tilde \Pi_{\nu} \\
\Deltaphi \\
\Vphitilde
\end{array}
\right)_{\mbox{neutrino iso.}} = C
\left(
\begin{array}{c}
3/4 \\
\Omega_{\gamma} \mathcal{P}\\
1 \\
\left(\Omega_{\gamma}+ \Omega_{\nu} + \frac{15}{4}\right) \mathcal{P}\\
3/4\\
- \Omega_{\gamma}/ \Omega_{\nu} \\
- \frac{15}{4} \,\mathcal{P}\,   \Omega_{\gamma} / \Omega_{\nu}\\
 - 3\,  \mathcal{P}\, \Omega_{\gamma}/\Omega_{\nu} \\
0\\
\Omega_{\gamma} \mathcal{P}
\end{array}
\right).
\end{equation}
It is important to note that we did not require quintessence to be
adiabatic. One can see from the neutrino isocurvature vector that
$\Deltaphi=0$, and as a consequence quintessence is not adiabatic with
respect to either neutrinos, radiation, baryons or CDM. 
Hence, we could just as well have labeled this vector ``quintessence
isocurvature''.  We cannot require adiabaticity between neutrinos, CDM, baryons
and radiation and hope to obtain a ``pure'' quintessence isocurvature
vector since, as we have seen in the discussion of the adiabatic mode,
these requirements lead to quintessence being adiabatic as well.


\subsection{\isocdm\ and \isobar}\label{sec::isocdm}
The CDM isocurvature mode is characterized by $S_{c:\gamma} \neq 0$, $\zeta_{tot}=0$
and adiabaticity between photons, neutrinos and baryons:
\begin{equation}\label{eqn::isocdmdel}
\zeta_{tot} = 0,\ \ \  \Delta_{\gamma} = \Dg{\nu} = \frac{4}{3} \Dg{b}.
\end{equation}
Using this and Equations (\ref{constraint_equation_start})-(\ref{constraint_equation_end}) yields
\ee
\bm{U}^T_{0}(\textrm{CDM iso.})= (1,0,0,0,0,0,0,0,0,0).
\eee
This vector fulfills $\zeta_{tot} =0 +\mathcal{O}(\omca)$, which is in line with our approximation since $\omca \ll 1$. 
Similarly, for the \isobar\ mode, we require $S_{b:\gamma} \neq 0$, $\zeta_{tot} =0$ and adiabicity
between photons, neutrinos and baryons. The resulting vector reads
\ee
\bm{U}^T_{0}(\textrm{baryon iso.})= (0,0,0,0,1,0,0,0,0,0).
\eee

As all but one of  the components of $\bm{U}_0$ are vanishing for \isocdm\ and 
\isobar, we use Equation (\ref{eqn::a1}) to obtain the next to constant order solution
for \isocdm
\newcommand{\uu}{\mathcal{U}}
\begin{equation} 
\left(
\begin{array}{c}
\Dg{c} \\
\tilde V_c\\
\Delta_{\gamma} \\
\tilde V_{\gamma} \\
\Dg{b} \\
\Delta_{\nu} \\
\tilde V_{\nu} \\
\tilde \Pi_{\nu} \\
\Deltaphi \\
\Vphitilde
\end{array}
\right)_{\mbox{CDM iso.}} = C
\left(
\begin{array}{c}
1 \\
 \omca(4 \omna-15) \, \uu /12 \\
0 \\
-(15/4)  \omca \, \uu   \\
0 \\
0 \\
 -(15/4)  \omca \, \uu   \\
- 2 \omca \, \uu      \\
  \omca (15+2 \omna) (1+ \wq) \, \uu \\
  \omca \, \uu \, \mathcal{V}
\end{array}
\right),
\end{equation}
where \mbox{$\uu = (30 +4 \omna)^{-1}$} and 
$\mathcal{V} = [105 -45 \wq +4\omna(3\wq-1)]/[36(\wq-1)]$.
Similarly, we find for \isobar
\begin{equation} 
\left(
\begin{array}{c}
\Dg{c} \\
\tilde V_c\\
\Delta_{\gamma} \\
\tilde V_{\gamma} \\
\Dg{b} \\
\Delta_{\nu} \\
\tilde V_{\nu} \\
\tilde \Pi_{\nu} \\
\Deltaphi \\
\Vphitilde
\end{array}
\right)_{\mbox{baryon iso.}} = C
\left(
\begin{array}{c}
0 \\
 \omba(4 \omna-15) \, \uu/12 \\
0 \\
-(15/4)  \omba \, \uu   \\
1 \\
0 \\
 -(15/4)  \omba \, \uu   \\
- 2 \omba \, \uu      \\
  \omba (15+2 \omna) (1+ \wq) \uu \\
  \omba \, \uu \, \mathcal{V}
\end{array}
\right).
\end{equation}

Note that these vectors are not constant since $\Omega_b$ and $\Omega_c$ both evolve in time.
We observe that the corrections to $U$ are indeed proportional  to $\omca$ or $\omba$ as expected. 
This result 
holds for all tracking quintessence models with $\wq=1/3$ or $\wq \leq 0$ during the radiation dominated 
period. 
For intermediate values $0 < \wq < 1/3$ the deviation from the leading behaviour 
scales $\propto x^{\alpha}$, $\alpha < 1$.
Obviously, the adiabatic, \isocdm, \isobar\ and neutrino isocurvature
vectors $\bm{U}_0$ are linearly independent. We have therefore identified 
four modes corresponding to the fourfold degenerate eigenvalue zero of $A(x)$.
These four vectors span the subspace of dominant modes in the
super-horizon limit, and there are no more linearly independent
vectors that satisfy the constraints (\ref{constraint_equation_start})
- (\ref{constraint_equation_end}). Arbitrary initial
perturbations may therefore be represented by projecting a
perturbation vector $\bm{U}$ at initial time into the
subspace spanned by the four aforementioned vectors, as this is the
part of the initial perturbations which will dominate as time progresses.


\begin{figure}[!t]
\begin{center}
\includegraphics[angle=0,scale=0.325]{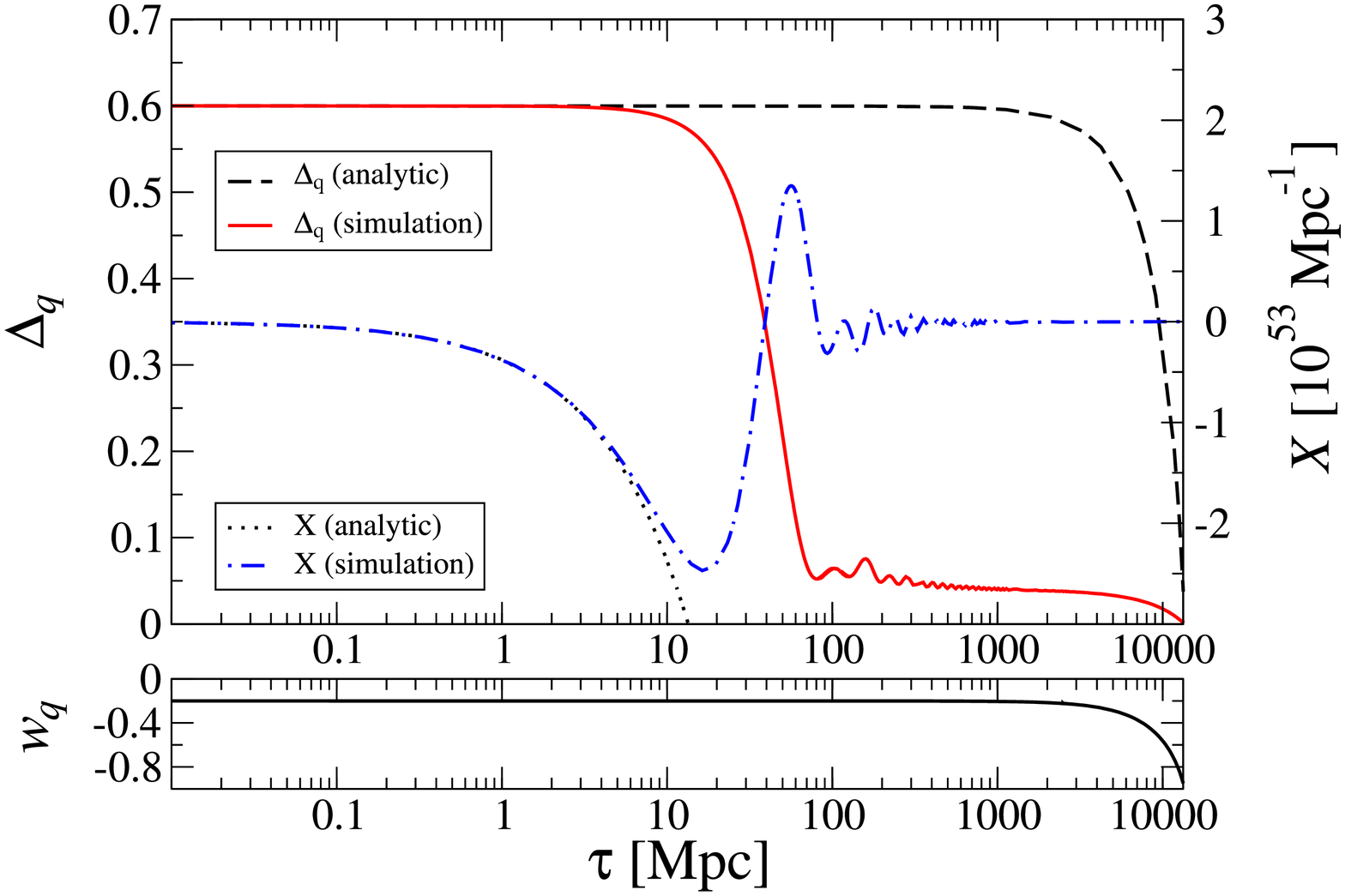}
\caption{Gauge-invariant energy density perturbation $\Deltaphi$ and  quintessence field fluctuation $X$ as simulated (straight
 and dashed-dotted lines),
compared to the analytic solution of Equations (\ref{adiabatic_mode}) and (\ref{eqn::Xana2}) (dashed and dotted lines) as a 
function of conformal time $\tau$ for 
adiabatic initial conditions. Radiation and matter equality corresponds to $\tau = 109 \rm Mpc$.
Shown is the mode for  $k = 0.1\,\mpc^{-1}$ and 
the cosmological parameters have been
$\omba^0 h^2 = 0.022,\ h=0.7,\ \Omega_{\rm m}^0 =  0.3,\ \Omega_{\rm q}^0=0.7$.}
\label{fig::iplX}
\end{center}
\end{figure}


Figure \ref{fig::iplX} demonstrates that the early time behaviour 
is well described by our analytic formulae.
The analytic results agree very well with the simulation for early times, when the mode is 
 outside the horizon. In the lower graph, we plot the equation of state $\wq$. The quintessence model 
used is parameterized 
by an equation of state $\wq(a) = -0.95 + 0.75(1-a)$, leading to $\wq({\rm early}) = -0.2$
and  according to (\ref{eqn::Xana2}), $X \propto \tau^{0.8}$.
 This differs from reference \cite{Perrotta:1998vf}. \footnote{In \cite{Perrotta:1998vf} it is stated that 
the quintessence fluctuation
  in Newtonian gauge scales $\propto \tau^2$ for  adiabatic initial conditions.
  This does not agree with our results in Appendix \ref{early}. Actually,
  equation (101) of \cite{Perrotta:1998vf} includes a factor
  $\varphi_{t0}$, which, interpreted as a dynamical quantity $\rd
  \varphi / \rd t$ (and not fixed at some initial time $t_0$), leads
  to a power law in $\tau$ which is then consistent with our result of
  Appendix \ref{early}.}

We see that including quintessence does not add a new dominant mode. The two additional modes added 
by the fluctuations of the scalar field are both subdominant and decay with negative 
eigenvalue $\lambda_i$. This is due to the fact that none of
the perturbation equations for quintessence equate to zero in the superhorizon limit. This holds for 
non-tracking quintessence models as well. Let us investigate this in detail. For all the other fluid 
components, $\Delta_a'=0$ in the super-horizon limit, but for quintessence we get 
from eq. (\ref{eq_dg}) that  
 $\Delta_q'=-3(c_{s(q)}^2-w_q)\Delta_q-3 w_q \Gamma_q$. For tracking quintessence, we obtain 
from equation (\ref{quint_sound}) that  $c_{s(q)}^2=w_q$ and we find
\begin{equation}
\Delta_q'=-3 w_q \Gamma_q
\end{equation}
Since $\Gamma_q$ does not vanish except for $w_q=1$ (see eq. (\ref{quint_gamma})), this does 
not equate to zero. \footnote{Note that $w_q=0$ does not lead to $\Delta_q'=0$.} Hence, due to the 
non-vanishing entropy perturbation of quintessence there 
is no additional dominant mode. \footnote{We have not yet investigated the relationship between 
decaying quintessence modes and the background evolution.}


\begin{figure*}[ht]
  \begin{center}
    \includegraphics[scale=0.5,angle=-90]{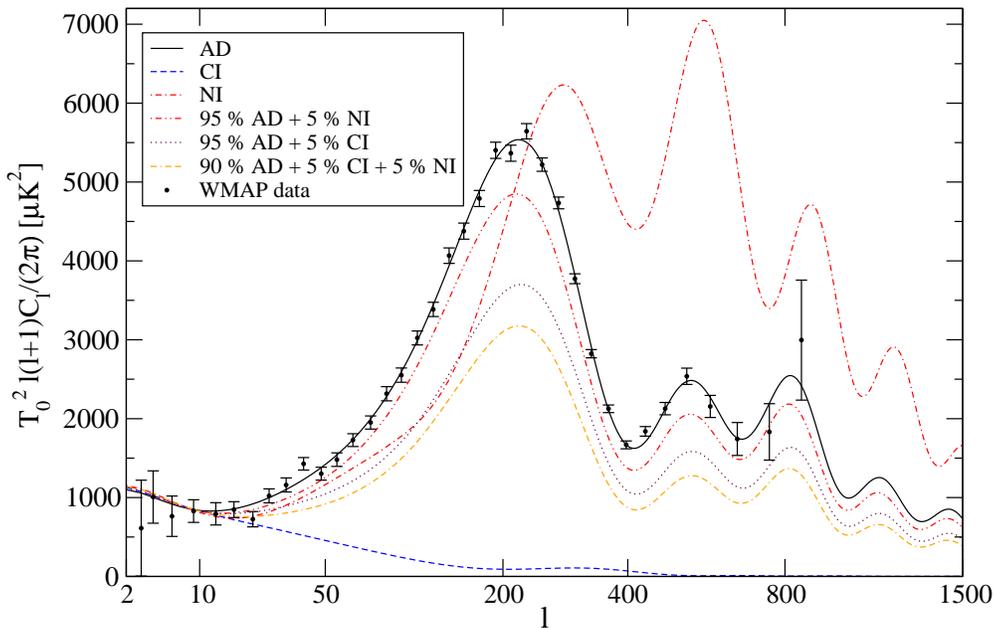}
    \caption{CMB Temperature spectra as a function of multipole $l$ in an early quintessence cosmology. 
      The pure adiabatic (AD), CDM isocurvature (CI), neutrino isocurvature (NI) mode and
  three different combinations of these dominant modes are plotted. For comparison with experimental data 
      we also give the WMAP measurements of the CMB \cite{Spergel:2003cb}. The spectrum of the 
pure \isobar\ mode is 
essentially identical to that of the pure \isocdm\ mode. All spectra have been normalized to the same 
power at $l=10$.}
    \label{example}
  \end{center}
\end{figure*}


\section{Isocurvature initial conditions and the CMB}  \label{section_example}

We illustrate the influence of different initial conditions on the CMB
with an example. For an analysis of experimental data and a
possible isocurvature contribution to the CMB we refer the reader to
\cite{Enqvist:2000hp,Trotta:2002iz,Gordon:2002gv}. Here, we merely whish to show the
qualitative features of the different modes. We use a modified version
of {\sc cmbeasy} \cite{Doran:2003sy,Seljak:1996is} to compute CMB spectra corresponding
to different initial conditions for an early quintessence cosmology
with parameters as in model A of \cite{Caldwell:2003vp}. We set the
spectral index of the isocurvature modes identical to the spectral
index of the pure adiabatic mode, $n_s = 0.99$. The resulting spectra
are plotted in Fig.\ \ref{example}.  The spectrum of the pure CDM
isocurvature mode decays quickly when going to small scales as has
been found in previous works \cite{Bucher:2000hy,Langlois:ar,Amendola:2001ni}. The
neutrino isocurvature mode shows prominent peaks at higher multipoles
than the adiabatic mode with different peak ratios. For the mixed
initial conditions with only small isocurvature contribution, the
shape of the curve remains more or less the same. A small admixture of isocurvature fluctuations leads
to a decrease of power at larger multipoles if the overall normalization is fixed at $l=10$. Comparison 
with the WMAP
data in the same figure shows that non-adiabatic initial perturbations
are strongly constrained. Clearly, pure isocurvature initial
conditions are inconsistent with CMB observations.


\section{Conclusion} \label{section_conclusion}
We have investigated perturbations in a radiation-dominated universe containing
quintessence, CDM, neutrinos, radiation and baryons in the tight
coupling limit. The perturbation evolution has been expressed as 
a differential equation involving a matrix acting on a vector comprised
of the perturbation variables.  This formulation leads to a systematic
determination of the initial conditions. In particular, we find that due to
the presence of tracking scalar quintessence no additional dominant mode is introduced. 
This fact is beautifully transparent in the matrix language. Indeed,
contributions of higher order in $x\equiv k\tau$ towards a perturbation vector $\bm{U}$
can easily be determined  by solving a simple matrix equation once the constant part of
$\bm{U}$ has been determined.

In total, we find four dominant modes and choose them as 
adiabatic, \isocdm, \isobar\ and \isonu. For the \isonu\ mode, quintessence
automatically is forced to non-adiabaticity. Hence, we could have as well labeled
the \isonu\ mode as quintessence isocurvature.
To demonstrate the influence on the cosmic microwave background anisotropy
spectrum, we have calculated spectra for all modes. Clearly, non-adiabatic contributions
are severely constrained by the data. 
A detailed study may provide ways to
put additional constraints on quintessence models or tell us more
about the initial perturbations after inflation.


\begin{acknowledgements}
We thank Robert R. Caldwell, Pier Stefano Corasaniti, Karim A. Malik and Roberto Trotta  
for helpful discussions. 
M. Doran was supported by NSF grant PHY-0099543 at Dartmouth and PHY-9907949 
at the KITP.  C.M. M{\"{u}}ller and G. Sch{\"{a}}fer are supported by GRK grant 216/3-02.
\end{acknowledgements}

\begin{appendix}


\section{Gauge-invariant perturbation equations}\label{section::gipe}
In this Appendix  we will explain the derivation of equations 
(\ref{eq_dcstrich})-(\ref{eq_poisson_sum}) in detail. 
\subsection{The general story}
First, we briefly summarize the gauge-invariant 
approach of \cite{Bardeen:kt,Kodama:bj,Durrer:2001gq}. 
Perturbing a homogenous Friedman universe, one
classifies fluctuations according to their transformation properties with respect to the rotation group.
In flat
spacetime, we may expand the perturbation variables in
terms of  harmonic
functions \cite{york}. With $Q_{,i}=\partial Q/\partial x^i$ one defines 
\begin{equation}
Q_{i}(\bm{k},\bm{x}) \equiv -k^{-1}Q(\bm{k},\bm{x})_{,i}
\end{equation}
and
\begin{equation}
Q_{ij}(\bm{k},\bm{x}) \equiv k^{-2}Q(\bm{k},\bm{x})_{,ij}+\frac{1}{3}\delta_{ij} Q(\bm{k},\bm{x}),
\end{equation}
where the $Q(\bm{k},\bm{x})$ are eigenfunctions of the Laplace-operator, $\nabla^2 Q_{k}(\bm{x})=-k^2 Q_{k}(\bm{x})$ and
in spatially flat universes $Q=\exp(i \bm{k}\bm{x})$. As modes with different $\bm{k}$ decouple
in linear theory, we will not display the $\bm{k}$-dependence of $Q$ in the following.
The scalar parts of vector and tensor fields can then be written as
\begin{equation}\label{eqn::bi}
B_i=BQ_i
\end{equation}
and
\begin{equation}\label{eqn::hij}
H_{ij}=H_LQ\delta_{ij}+H_TQ_{ij}.
\end{equation}
respectively.

 In this work, we are only interested in  scalar fluctuations
because  scalar quintessence will not influence vector or tensor modes. 
The most general line element for a perturbed
Robertson-Walker metric may  be written as
\begin{eqnarray}\label{eq_metric} \nonumber
\textrm{d}s^2=a(\tau)^{2} [-(1+2A) \textrm{d} \tau^{2}-2B_{i} \textrm{d}\tau \textrm{d}x^i+ \\ 
(\delta_{ij}+2H_{ij}) \textrm{d}x^{i} \textrm{d}x^{j}],
\end{eqnarray}
where in the scalar case $B_i$ and $H_{ij}$ are given by equations (\ref{eqn::bi}) and (\ref{eqn::hij}).
The gauge transformation of a tensor $T$ is given by \cite{Bardeen:kt,Kodama:bj,Mukhanov:1990me,Durrer:2001gq,Doran:2003sy}
\begin{equation}
\tilde{T}(x) = T(x) - L_{\epsilon} \bar{T},
\end{equation}
where $L_{\epsilon}$ is the Lie derivative. The transformation vector $\epsilon$ can be decomposed as 
\begin{equation}
\tilde{\tau}= \tau + T(\tau) Q(\bm{x}),
\end{equation}
\begin{equation}
\tilde{x}^{i}= x^{i} + L(\tau)Q^{i}(\bm{x}),
\end{equation}
where $L$ and $T$ are arbitrary functions of $\tau$. The transformation properties of the metric perturbations are given by
\cite{Bardeen:kt,Doran:2003sy}
\begin{eqnarray}\label{eq_a}
\tilde{A}&=&A-\hubble T-\dot{T}, \\ \label{eq_b}
\tilde{B}&=&B+\dot{L}+kT, \\ \label{eq_hl}
\tilde{H}_{L}&=&H_L-\hubble T-\frac{k}{3}L, \\ \label{eq_ht}
\tilde{H}_{T}&=&H_T+kL,
\end{eqnarray}
where a dot denotes the derivative with respect to conformal time
$\tau$ and $\hubble \equiv \dot{a}(\tau)/a(\tau)$.  The functions $L$ and $T$ can be used to eliminate two
of the metric perturbations. Popular choices are
$A=B=0$ for the synchronous gauge and
$B=H_T=0$ for the longitudinal gauge.

 From equations
(\ref{eq_a})-(\ref{eq_ht}) one can construct 
the gauge-invariant Bardeen potentials \cite{Bardeen:kt}
\ea\label{eq_psi}
\Phiflex=A-\hubble k^{-1}\sigma-k^{-1}\dot{\sigma},\\\label{eq_phi}
\Psiflex=H_L+\frac{1}{3}H_T-\hubble k^{-1}\sigma,
\eea
with $\sigma \equiv k^{-1} \dot{H}_{T} -B$.
It is worthwhile to note that in longitudinal gauge, for which $B=H_T=\sigma=0$,
the perturbed metric takes on the simple form
\begin{multline}\label{eqn::longitudinal}
ds^2_{(long.)}  = a(\tau)^2 \big [ -(1 + 2\Phiflex\, Q) d\tau^2 \\ + ( 1  + 2 \Psiflex\,Q) \delta_{ij}dx^i dx^j \big ].
\end{multline}
With $\Mp \equiv (8 \pi G)^{-1/2}$ denoting the reduced Planck mass, 
Einstein's equation reads
\begin{equation} \label{eq_einst}
T^{\mu}_{\;\:\nu} = \Mp^2 \left( R^{\mu}_{\;\, \nu} - \frac{1}{2} \delta^{\mu}_{\; \nu} R \right),
\end{equation}
where the energy momentum tensor of a perfect fluid is given by 
\begin{equation}
T^{\mu}_{\;\:\nu} = p \delta^{\mu}_{\;\nu} + (\rho +p)u^{\mu} u_{\nu} +\pi^{\mu}_{\;\,\nu}.
\end{equation}
The covariant 4-velocity is $u_i =a (v(\tau)-B)\, Q_i$. We define the
energy density contrast $\delta$ by $\rho= \bar{\rho}\,
(1+\delta(\tau)) Q$, the spatial trace by $p \, \delta_{\,j}^i=
\bar{p}(\tau)(1+\pi_L (\tau) Q) \,\delta_{\,j}^i$ and the traceless
part by $\pi_{\;j}^i = \bar{p}\; \Pi \;Q_{\:j}^i$. Therefore
the components of the energy momentum tensor are
\begin{eqnarray}
T_{\;\;0}^0&=& -\bar{\rho} (1+ \delta Q), \\
T^i_{\;\;0}&=& -\bar{\rho} (1+w)\, v\, Q^i,  \\
T^0_{\;\;i}&=& \bar{\rho} \,(1+w)\, (v-B)\, Q_i, \\
T^i_{\;\;j}&=&\bar{p} \,\left[\left(1+\pi_LQ \right) \delta^i_{\,j} + \Pi Q^i_{\,j} \right] \label{eqn::defPi}.
\end{eqnarray}
Given the gauge-transformation properties of  $\delta$, $v$ and $\pi_L$ 
\cite{Bardeen:kt,Kodama:bj,Mukhanov:1990me,Durrer:2001gq,Doran:2003sy}, one can construct 
gauge-invariant quantities
for the energy density contrast $\Delta$, the 
velocity $V$ and the entropy perturbation $\Gamma$.  These are given by
\begin{eqnarray}
\Delta&=&\delta +3(1+w)\left(H_L +\frac{1}{3} H_T\right), \label{eqn::defD} \\ 
V&=&v-k^{-1} \dot{H}_{T}, \label{eqn::defV}\\
\Gamma &=& \pi_L -\frac{c_s^2}{w} \delta.
\end{eqnarray}
Here, $c_s^2 \equiv \partial \bar p / \partial \bar \rho$ is the adiabatic sound speed.
From the conservation of the zero component of the energy momentum tensor $\nabla_{\mu}\bar{T}_{\;\;0}^{\mu}=0$ we obtain   
\begin{equation} \label{eq_rho}
\frac{\dot{\bar{\rho_{\alpha}}}}{\bar{\rho_{\alpha}}}=-3(1+w_{\alpha})\,\hubble ,
\end{equation}
where $w=\bar{p}/\bar{\rho}$ is the equation of state of the
particular species.  The perturbed Einstein equations in gauge-invariant
variables are \cite{Bardeen:kt,Kodama:bj,Mukhanov:1990me,Durrer:2001gq,Doran:2003sy}:
\begin{eqnarray} \nonumber
a^2  \bar\rho \Delta &=&  2\,\Mp^2\, k^2 \Psiflex  \\   &&-   3 a^2 \bar{\rho}\,(1+w) \left(\hubble  k^{-1}V -\Psiflex \right)\!, 
\label{eq_eing00}  \\
a^2   (\bar\rho  + \bar p ) V &=&  2\,\Mp^2\, k\left(\hubble \Phiflex - \dot \Psiflex\right) \label{eq_eingi0}, \\
a^2  \bar p\,\Pi &=& - \Mp^2 \,k^2(\Psiflex + \Phiflex),  \label{eq_eingij}
\end{eqnarray}
In the above, it is understood that the quantities $\Delta,\ V$ and $\Pi$ are the
sum of the contributions of all species $\alpha$. 
 Using $
\dot{w}= (c_s^2-w) \dot{\bar{\rho}}/\bar{\rho} $ and (\ref{eq_rho}) we
get from $T^{\mu}_{\;\;0;\mu}=0$ that
\begin{equation}\label{eq_dg}
\dot{\Delta}+3(c_s^2-w)\hubble  \Delta +kV(1+w)+3\hubble w\Gamma=0,
\end{equation}
and from $T^{\mu}_{\;\:i;\mu}=0$ 
\begin{eqnarray}\label{eq_v}  \nonumber 
\dot{V}=\hubble (3c_s^2-1)V+k[\Phiflex-3c_s^2\Psiflex]
\\ 
+\frac{c_s^2k}{1+w} \Delta +\frac{wk}{1+w} \Bigg[ \Gamma-\frac{2}{3}\Pi \Bigg].
\end{eqnarray}

\subsection{Gauge-invariant quintessence perturbations}
The scalar quintessence field is decomposed into a background
and fluctuation part according to $\varphi(\tau, \bm{x}) =
\bar{\varphi}(\tau) + \chi(\tau, \bm{x})$. The fluctuation can be 
promoted to a gauge-invariant quantity by defining the
gauge-invariant quintessence field fluctuation 
$X\equiv \chi - \dot{\bar{\varphi}} k^{-1}\sigma$.
The field dynamics is governed by the Klein-Gordon equation. 
For the background, it reads:  
\begin{equation}\label{eq_kgback}
\ddot{\bar{\varphi}} = -2 \hubble  \dot{\bar{\varphi}} - a^2 V'(\varphi),
\end{equation}
while the perturbation obeys the equation of motion
\begin{multline}\label{eq_kgpert}
\ddot{X} = \dot{\bar{\varphi}}(\dot{\Phiflex} -3\dot{\Psiflex})-2a^2 V'(\varphi) \Phiflex \\ -(a^2  V''(\varphi)+k^2)X-2 
\hubble  \dot{X}.
\end{multline}


\begin{table}[t]
\begin{ruledtabular}
\begin{tabular}{ccc}
Symbol &  Meaning & Equation \\
\hline
$\Omega_{\rm species}$ &  fraction of  total energy density & n.a.\\ 
$\Omega_{\rm species}^0$ &  fraction of  total energy density today & n.a.\\ 
$a$ & scale factor of the universe & n.a. \\
$\tau$ & conformal time: $\rd \tau = \rd t  / a$ & n.a. \\
$k$ & wavenumber of mode & n.a. \\
$x$ & $k \tau$ & n.a. \\
$\dot{}$ & derivative w.r.t conformal time & n.a. \\
$'$ & derivative w.r.t.  $x \frac{\rd}{\rd x}$ & n.a. \\
$\hubble$ & $\dot a / a$ & n.a. \\
\hline
$\Delta$ & gauge-inv. density contrast ($\Delta_g$ of \cite{Kodama:bj}) & (\ref{eqn::defD}) \\
$V$ & gauge-invariant velocity &  (\ref{eqn::defV}) \\
$\Pi$ &  shear & (\ref{eqn::defPi}) \\
$\tilde V$ & reduced velocity: $\tilde V = x^{-1} V$ & n.a. \\
$\tilde \Pi$ & reduced shear: $\tilde \Pi = x^{-2} \Pi$ & n.a. \\
\end{tabular}
\end{ruledtabular}
\caption{Symbols and their meanings.}

\end{table}


From the energy momentum tensor for the quintessence field
\begin{equation}
T^{\mu}_{\;\,\nu}=\varphi^{,\mu} \varphi_{,\nu} - \delta^{\mu}_{\;\nu} \left( \frac{1}{2}\varphi^{,\alpha} \varphi_{,\alpha}+V(\varphi) \right),
\end{equation}
using $\varphi = \bar\varphi + X$  and the longitudinal gauge metric, one gets
\begin{eqnarray}
\hspace{-2em}\delta T^{0\ (lon.)}_{\ 0} &=& \left[a^{-2}\left( \dot{\bar\varphi}^2\, \Psiflex - \dot X \dot{\bar\varphi} \right) - V^{\prime}(\varphi) X\right] Q \label{eqn::deltaTq1} \\
\hspace{-2em}\delta T^{i\ (lon.)}_{\ 0} &=& -a^{-2}\,k\,  \dot{\bar\varphi}\,X \,Q^i \label{eqn::deltaTq4}.
\end{eqnarray}
Using the definition of $\Delta$,  equation (\ref{eqn::defD}) in longitudinal gauge
and  $\bar{\rho}_q+\bar{p}_q = a^{-2}\dotphi^2$  one
can read off from equation (\ref{eqn::deltaTq1})
the gauge-invariant expression
\begin{equation}\label{eq_dphi1}
\Deltaphi= (1+ \wphi) \left[ 3\Psiflex -\Phiflex +\dot{X} \dot{\bar{\varphi}}^{-1} \right] +X V'(\varphi) \bar{\rho}_q^{(-1)}.
\end{equation}
In the same manner, one gets from equation (\ref{eqn::deltaTq4}) and the
fact that $v^{(long.)} = V$ the relation
\begin{equation}\label{eq_vphi1}
\Vphi = k \dot{\bar{\varphi}}^{-1} X.
\end{equation}
Taking the time derivative of equations (\ref{eq_dphi1}) and (\ref{eq_vphi1}) and
using the equation of motion  (\ref{eq_kgpert}), one obtains the evolution equations
\begin{eqnarray}\label{eq_dphi}  \nonumber
\Deltaphidot = (1 &+& \wphi) \Bigg[ \frac{2a^2 V'(\varphi)}{\dot{\bar{\varphi}}} \Bigg( \frac{\Deltaphi}{1+ \wphi} - 3\Psiflex \Bigg)
\\ 
&+& \Bigg( \frac{6a\dot{a} V'(\varphi)}{k \dot{\bar{\varphi}}}-k \Bigg) \Vphi \Bigg] + \frac{\dot{\wphi} \Deltaphi}{1+ \wphi}
\end{eqnarray}
and
\begin{equation}\label{eq_vphi}
\Vphidot = k \left[ \frac{\Deltaphi}{1+ \wphi}- 3\Psiflex +\Phiflex \right] +2 \hubble   \Vphi.
\end{equation}

Equation (\ref{eq_dphi}) depends on the specific quintessence model
through $V'$ and $\dotphi$. We can however make progress in the case of nearly
constant $\wq$:   
Many quintessence models have solutions for which $\varphi$ approaches
an attractor solution irrespectively of its initial value. 
For these
tracking quintessence models \cite{Wetterich:fm,Ratra:1987rm,Steinhardt:nw}, the
equation of state of the quintessence field $\wphi$ is nearly constant
during radiation domination. We will use this vanishing of $\dot \wphi$ in the
following to derive relations to simplify equation (\ref{eq_dphi}).
Considering 
$a^{-2}\dotphi^2  = (1+\wphi)  \rho_\varphi$
it follows using the Friedman equation $3a^{-2}\Mp^2\hubble^2 = \rho$ that
\begin{equation} \label{eq_phiscale}
\dot{\bar{\varphi}}= [3(1+ \wphi) \Omegaphi]^{\frac{1}{2}} \Mp \hubble,
\end{equation}
and hence
\ee\label{eqn::dotphi}
\frac{\ddotphi}{\dotphi} = \frac{\rd}{\rd \tau} \ln \dotphi = \frac{1}{2} \frac{\dot \omqa}{\omqa} + \frac{\dot\hubble}{\hubble},
\eee
where we have neglected a term involving $\dot \wphi$. 
We will in the following assume
that at early times, the universe expands as if radiation dominated.
In this case, $\hubble = \tau^{-1}$ and  
inserting the above equation (\ref{eqn::dotphi}) 
into the equation of motion (\ref{eq_kgback}), one finds
\ee\label{eqn::simple}
\frac{a^2 V'}{\dotphi} = -\frac{3(1-\wq)}{2\, \tau}.
\eee
Using this relation (\ref{eqn::simple}), the evolution equation for $\Deltaphi$ becomes
\begin{multline}
\Deltaphidot = 3 (\wq-1) \frac{k}{x} \bigg[  \Deltaphi - 3(1+\wq) \Psiflex 
\\ + \left \{3 - \frac{x^2}{3(\wq -1)} \right \}\, (1+\wq) \Vphitilde \bigg ],
\end{multline}
whereas the one for the velocity remains almost unaltered while we move
to the reduced velocity $\Vphitilde$:
\ee
\Vphitildedot = \frac{k}{x} \left[ \frac{\Deltaphi}{ 1+ \wphi} -3\Psiflex + \Phiflex \right] + \tau^{-1}  \Vphitilde.
\eee
Note that $\Gamma_q$ does not usually vanish. Instead, we obtain
\begin{equation}\label{quint_gamma}
w_q \Gamma_q=(1-c_{s(q)}^2)\left[\Delta_{q}-3(1+w_q)\Phi+3\frac{\dot a}{a} (1+w_q)\frac{V_q}{k}\right]
\end{equation}
with the sound speed of quintessence given by
\begin{equation}\label{quint_sound}
c_{s(q)}^2=\dot{\overline{p}}_q/\dot{\overline{\rho}}_q = w_q-\frac{1}{3} \frac{a}{\dot a} \frac{\dot w_q}{1+w_q}
\end{equation}


\subsection{Matter and Radiation}
Setting $w=c_s^2=\Gamma=0$ in equations (\ref{eq_dg}) and (\ref{eq_v}), we
obtain the cold dark matter evolution equations 
\ea
\dot{\Delta}_{\rm c} &=&-kx \tilde{V}_{\rm c}, \label{eq_dc}  \\ 
\dot{\tilde{V}}_{\rm c} &=& \frac{k}{x}  ( -\tilde{V}_{\rm c} +\Phiflex ). \label{eq_vc}
\eea
The multipole expansion of the 
neutrino distribution function \cite{Ma:1995ey,Durrer::fund} can be truncated beyond the quadrupole at early
times. In terms of density, velocity
and shear, it is given by \cite{Durrer::fund,Doran:2003sy}
\ea
\dot{\Delta}_{\nu}&=& - \frac{4}{3}  k x \tilde{V}_{\nu},  \label{eq_dnu} \\ \label{eq_vnu}
\dot{\tilde{V}}_{\nu}&=& \frac{k}{x} \left(  \frac{1}{4} \Delta_{\nu} -\tilde{V}_{\nu} -\frac{1}{6}  x^2 \tilde{\Pi}_{\nu} + 
\Phiflex- \Psiflex \right), \\ \label{eq_pinu}
\dot{\tilde{\Pi}}_{\nu} &=& \frac{k}{x} \left( \frac{8}{5} \tilde{V}_{\nu} -2 \tilde{\Pi}_{\nu} \right).
\eea
Deep in the radiation dominated era, for which the initial conditions
here are derived, Compton scattering tightly couples photons and
baryons \cite{Peebles:ag,Kodama:bj}. The coupling leads 
to $V_{\rm b} = V_\gamma$ and the evolution equations 
become \cite{Kodama:bj}
\ea
\dot{\Delta}_{\gamma}&=& - \frac{4}{3} k x \tilde{V}_{\gamma}, \label{eq_dgamma} \\ \label{eq_vgamma}
\dot{\tilde{V}}_{\gamma}&=& \frac{k}{x} \left( \frac{1}{4} \Delta_{\gamma} -\tilde{V}_{\gamma} + \Phiflex-\Psiflex \right), \\
\dot{\Delta}_{\rm b} &=& -k\, x\,  \tilde V_\gamma  \label{eqn_dgbaryon}.
\eea 
As the photon quadrupole and all higher photon multipoles are suppressed during 
tight coupling, it follows that $\Psiflex$ is given from Einstein's equation by
\ee\label{eqn::phipsiphi}
\Psiflex = -\Phiflex  -  \Omega_\nu \tilde{\Pi}_{\nu},
\eee
where we have used the Friedmann equation.
Finally, the Poisson equation (\ref{eq_eing00}) in terms of the various species 
is 
\ee
 \label{eq_poisson_sum2}
\Phiflex=- \frac{ \sum\limits_{\alpha=c,b,\gamma,\nu,q} \Omega_{\alpha} ( \Delta_{\alpha} 
+3(1+w_{\alpha}) \tilde V_{\alpha})}{ \sum\limits_{\alpha=c,b,\gamma,\nu,q}  
3(1+w_{\alpha})\Omega_{\alpha} + \frac{2 x^{2}}{3}} - \Omega_{\nu} \tilde{\Pi}_{\nu},
\eee
where
the index $\alpha$ runs over all species.
Rewriting the evolution equations (\ref{eq_dc}) - (\ref{eqn_dgbaryon}) in terms
of $\rd / \rd \ln x$ and replacing $\Psiflex$ by means of   (\ref{eqn::phipsiphi}),
one arrives at (\ref{eq_dcstrich})-(\ref{eq_q2}).


\begin{table}[t]
\begin{ruledtabular}
\begin{tabular}{cl}
Quantity &  Scaling behaviour\\
\hline
$\dotphi$ &   $\propto \tau^{-(1+3\wq)/2}$ \\
$V'$ & $\propto \tau^{-(7+3\wq)/2}$ \\
$V''$& $\propto \tau^{-4}$  \\
$\Deltaphi^{\rm adiab.}$ & $const.$  \\
$X_{\rm adiab.}$ & $\propto \tau^{(1-3\wq)/2}$
\end{tabular}
\end{ruledtabular}
\caption{Tracking quintessence in the radiation era: Scaling handbook.}
\end{table}


\section{Early time quintessence field fluctuations}\label{early}
While throughout this work, we describe quintessence perturbations
by the variables $\{\Deltaphi,\Vphi \}$, one could instead use the field
fluctuation and its time derivative $\{X,\dot X\}$. In this
section, we will give analytic expressions for $X$ and $\dot X$ in
the case of tracking quintessence for super-horizon modes. 
We will do so assuming that $\Phiflex$ and $\Psiflex$ are at least almost constant.
As this is not the case for \isocdm\ and \isobar, the following steps
do not apply in these modes. Furthermore, we will assume
that the universe expands as if radiation dominated during the
time of interest. In this case, $\hubble = \tau^{-1}$, $\Omegaphi \propto \tau^{1-3\wq}$ and hence by
means of  equation (\ref{eq_phiscale}) 
$\dotphi \propto \tau^{-\frac{1}{2}(1+3\wq)}$. Using this, 
we infer from 
equation  (\ref{eqn::simple}) that $V' \propto  \tau^{-\frac{1}{2}(7+3\wq)}$.
In addition, a straightforward calculation using (\ref{eqn::dotphi}) and (\ref{eqn::simple})
yields 
\ee\label{eqn::vprime2}
a^2 \tau^2 V'' = a^2 \tau^2  \frac{\rd V'}{\rd \tau}\frac{\rd \tau}{\rd \varphi} = \frac{3}{4} (1-\wq)(7+3\wq).
\eee
The equation of motion for $X$ 
(\ref{eq_kgpert}) contains a term
$\dotphi\left(\dot\Phiflex-3\dot\Psiflex\right)$, which by assumption we may drop.
In addition, we see from equation (\ref{eqn::vprime2}), that 
for super-horizon modes, $a^2V^{\prime\prime} \gg k^2$ (except for $\wq$ very close to $1$),
and hence the equation of motion reduces to 
\ee\label{eqn::eomearly}
\ddot X = -2 a^2 V^{\prime} \Phiflex - a^2 V^{\prime\prime} X - 2\adota \dot X.
\eee
Using the power law behaviour in $\tau$ of 
$V',\ V''$ and $a$, as well as equations (\ref{eqn::simple}) (\ref{eqn::vprime2}), one finds 
the particular solution
\ee\label{eqn::Xana2}
X(\tau) = \frac{\tau}{2} \Phiflex\, \dotphi,
\eee
as well as two complementary solutions that may be added to obtain
the general solution
\begin{multline}\label{eqn::generalsolu}
X(\tau) = \frac{\tau}{2} \Phiflex\, \dotphi + 
c_1 \,\tau^{-\frac{1}{2}\left(1-\sqrt{1-4 a^2 \tau^2 V'' }\right)} \\+ 
c_2\,\tau^{-\frac{1}{2}\left(1+\sqrt{1-4 a^2 \tau^2 V''} \right)}.
\end{multline}
The mode  proportional to $c_2$ is at least as rapidly decaying 
as the one proportional to $c_1$. Using the explicit form of $4 a^2 \tau^2 V''$,
equation (\ref{eqn::vprime2}), we find that $\sqrt{1-4 a^2 \tau^2 V''}$
is imaginary if $\wq \in [-\frac{2}{3}(1+\sqrt{6}), -\frac{2}{3}(1-\sqrt{6})]$,
which holds for all scalar quintessence models of current interest.
Hence, the complementary modes decay $\propto 1/\sqrt{\tau}$ in an oscillating
manner.

Coming back to the dominating particular solution
(\ref{eqn::Xana2}), Figure \ref{fig::iplX} shows that the accuracy of
this analytic result is indeed high at early times, when compared to
numerical simulations.

Inserting the solution (\ref{eqn::Xana2}) and its time derivative into
equation (\ref{eq_dphi1}), we find the simple expression
\ee
\Deltaphi = 3(1+\wq) \left (\Psiflex - \frac{1}{2}\Phiflex \right),
\eee
which is just a restatement of eqn. (\ref{constraint_equation_onebeforeend}) and (\ref{constraint_equation_end}).
Hence, the energy density contrast in tracking quintessence 
models remains constant on super horizon scales, provided the gravitational
potentials are constant to good approximation.


\section{Extended Matrix Formulation}\label{sec::extend}
For simplicity, we have limited the discussion in section (\ref{sec::matrix}) to
cases where either $\wq = 1/3$, or quintessence contributions to $A(x)$ are
neglected. Here, we will discuss cases for which $\wq < 1/3$, while the
background expands radiation dominated. In this case, $\omqa \propto \tau^{(1-3\wq)}$
and we can split the matrix in three parts according to their scaling with $x$:

\ee
A(x) = A_0 + x\, A_1 + x^{(1-3 \wq)}\, A_q.
\eee
Again, Equation (\ref{eqn::matrix}) will lead to a solution vector of the form
\ee
\bm{U}(x) = \bm{U}_0 + x\, \bm{U}_1 +  x^{(1-3 \wq)}\,\bm{U}_q.
\eee
Substituting this into Equation (\ref{eqn::matrix}) and keeping only leading orders in $x$,
we get
\ea
A_0 \,\bm{U}_0 &=& 0, \\
A_1 \,\bm{U}_0 + A_0 \,\bm{U}_1 &=& \bm{U}_1, \label{eqn::rawa1} \\
A_q \,\bm{U}_0 + A_0 \,\bm{U}_q &=& (1-3\wq) \,\bm{U}_q.\label{eqn::rawaq}
\eea
While the conclusion regarding $\bm{U}_0$ and $\bm{U}_1$ are still the same as
in section (\ref{sec::matrix}), we see that quintessence may introduce a correction
\ee
\bm{U}_q = -\left[A_0 - (1-3\wq) \openone \right]^{-1} A_q \,\bm{U}_0.
\eee
This contribution $x^{(1-3\wq)} \,\bm{U}_q$ could in principle 
dominate over $x \,\bm{U}_1$ for $\wq > 0,\, \omqa > \omca$. 
However, the contribution is only of interest for the \isocdm\ and \isobar\ modes,
as it is otherwise negligible compared to the constant order. Yet for \isocdm\
and \isobar, $A_q\, \bm{U}_0=0$. Therefore, the discussion below applies, leading
to $\bm{U}_q=0$ for \isocdm\ and \isobar\ modes.  One
order higher in $x$, there may be a contribution. Yet this is in any case a 
higher order contribution, which we may neglect. 

Finally, we briefly discuss the case of vanishing $\bm{U}_0$. This only concerns possible subdominant modes. Equation
(\ref{eqn::rawa1}) then yields 
$A_0 \,\bm{U}_1 = \bm{U}_1$, i.e. $\bm{U}_1$ is an eigenvector of $A_0$ with eigenvalue $\lambda=1$.
As $A_0$ does not have such an eigenvector, we are led to conclude that 
Equation (\ref{eqn::matrix}) does not have a regular solution involving $\bm{U}_1$, if $\bm{U}_0=0$.
Turning to Equation (\ref{eqn::rawaq}), we similarly conclude that
$\bm{U}_q$ needs to be a eigenvector of $A_0$ with $\lambda=(1-3\wq)$ for vanishing $\bm{U}_0$.
For $\wq < 1/3$ this is once again excluded and  for $\wq = 1/3$, we just regain 
the results of section \ref{sec::matrix}.

\end{appendix}


\bibliographystyle{unsrt}

\end{document}